# Purifying selection, drift and reversible mutation with arbitrarily high mutation rates

**Authors:** Brian Charlesworth[*] and Kavita Jain[§]

**Affiliation:**

[*]*Institute of Evolutionary Biology, School of Biological Sciences, University of Edinburgh, Edinburgh EH9 3JT, United Kingdom*

[§]*Theoretical Sciences Unit, Jawaharlal Nehru Centre for Advanced Scientific Research, Jakkur P.O., Bangalore 560064, India*

*Running head*: Population genetics of hyperdiversity

*Corresponding Author*:

Brian Charlesworth

Institute of Evolutionary Biology, School of Biological Sciences

University of Edinburgh

Ashworth Laboratories

King's Buildings

Edinburgh EH9 3JT, UK

*Telephone*: +44 (0)131 650 5751

*Fax*: +44 (0)131 650 6564

*Email*: Brian.Charlesworth@ed.ac.uk

**ABSTRACT** Some species exhibit very high levels of DNA sequence variability; there is also evidence for the existence of heritable epigenetic variants that experience state changes at a much higher rate than sequence variants. In both cases, the resulting high diversity levels within a population (hyperdiversity) mean that standard population genetics methods are not trustworthy. We analyze a population genetics model that incorporates purifying selection, reversible mutations and genetic drift, assuming a stationary population size. We derive analytical results for both population parameters and sample statistics, and discuss their implications for studies of natural genetic and epigenetic variation. In particular, we find that (1) many more intermediate frequency variants are expected than under standard models, even with moderately strong purifying selection (2) rates of evolution under purifying selection may be close to, or even exceed, neutral rates. These findings are related to empirical studies of sequence and epigenetic variation.

The infinite sites model, originally proposed by Fisher (1922,1930) and developed in detail by Kimura (1971), has been the workhorse of molecular population genetics for four decades. Its core assumption is that any nucleotide site segregates for at most two variants, and that the mutation rate scaled by effective population size ($N_e$) is so low that new mutations arise only at sites that are fixed within the population. This assumption facilitates calculations of the theoretical values of some key observable quantities, such as the expected level of pairwise nucleotide site diversity or the expected number of segregating sites in a sample (Kimura 1971; Watterson 1975; Ewens 2004). In the framework of coalescent theory, this implies a linear relation between the genealogical distance between two sequences and the neutral sequence divergence between them, greatly simplifying methods of inference and statistical testing (Hudson 1990; Wakeley 2008).

There has recently been some discussion of how to go beyond the infinite sites assumption of a low scaled mutation rate, which breaks down for species with very large effective population sizes, including some species of virus and bacteria, and even eukaryotes such as the sea squirt and outbreeding nematode worms, resulting in "hyperdiversity" of DNA sequence variability within a population (Cutter *et al.* 2013). It is important to note, however, that this problem can arise even when the scaled mutation rate is relatively low, since there the proportion of neutral nucleotide sites that are currently segregating in a population (which depends on the scaled mutation rate) can be substantial when the population size is sufficiently large. For example, with a neutral mutation rate of $u$ per site in a population of $N$ breeding adults, the expected fraction of sites that are segregating in a randomly mating population is $f_s = \theta [\ln (2N) + 0.6775]$, where $\theta = 4N_e u$ (Ewens 2004, p.298). Thus, with $\theta = 0.01$, a reasonable value for many species (Leffler *et al.* 2012), we have $f_s = 0.15$ even when $N$ has the implausibly low value of one million. This implies that about 15% of new mutations are expected to arise at sites that are already segregating, suggesting a significant departure from the assumptions of the infinite sites model. (An alternative way of looking at this is to determine the expected number of new mutations that occur at a site while a pre-existing mutation is segregating, which is of a similar magnitude to $f_s$ – see Appendix, equation (A1).)

In addition, it has been known for nearly twenty years that sufficiently high scaled mutation rates at some or all sites in a sequence can lead to substantial departures from the infinite sites expectations for statistics such as Tajima's $D$, which are commonly used to detect deviations from neutral equilibrium caused by population size changes or selection (Bertorelle and Slatkin 1995; Aris-Brosou and Excoffier 1996; Tajima 1996; Yang 1996; Mizawa and Tajima 1997). This is because the occurrence of mutations at sites that are already segregating increases the pairwise diversity among sequences, but does not increase the number of segregating sites (Bertorelle and Slatkin 1995). The analysis of data on DNA sequence variation in hyperdiverse species thus requires methods that deal with this problem, and a number of population genetics models that contribute to this have already been developed (Desai and Plotkin 2008; Jenkins and Song 2011; Cutter *et al.* 2012; Jenkins *et al.* 2014; Sargsyan 2014).

Finally, analyses of the inheritance of epigenetic markers, such as methylated cytosines, have suggested that these can sometimes be transmitted across several sexual generations, but with rates of origination or reversion that are several orders of magnitude higher than the mutation rates of DNA sequences (Johannes *et al.* 2009; Becker *et al.* 2011; Schmitz *et al.* 2011; Lauria *et al.* 2014). In view of the current interest in the possible functional and evolutionary significance of epigenetic variation (Richards 2006; Schmitz and Ecker 2012; Grossniklaus *et al.* 2013; Klironomos *et al.* 2013), it seems important to develop models that can shed light on their population genetics, in order to understand the evolutionary forces acting on them.

The purpose of the present paper is to develop a relatively simple analytical framework for examining the consequences of high scaled mutation rates, in the framework of the classical random mating, finite population size model with forward and backward mutations in the presence of selection and genetic drift (Wright 1931; Wright 1937). The approach is similar in spirit to the biallelic model used by Desai and Plotkin (2008), but with a focus on sample statistics that summarize properties of the site frequency spectrum, as well as on the expected rate of substitutions along a lineage. As has been found in previous coalescent-based treatments with neutrality (Bertorelle and Slatkin 1995; Aris-Brosou and Excoffier 1996; Cutter *et al.* 2012), the results derived below show that very large departures from the infinite sites model occur when the scaled mutation rate is sufficiently high, even when fairly strong purifying selection is acting, resulting in features of the data such as a large excess of intermediate frequency variants. In addition, the signal of purifying selection on substitutions along a lineage can be obscured, or even converted into a signal of positive selection. The findings have significant implications for the interpretation of the results of studies of both epigenetic variability and DNA sequence variability in species with large effective population sizes.

## Analysis of the Model of Purifying Selection, Drift and Mutation

### Basic assumptions

In order to generate simple analytical results, we use a "finite-sites" model that is an extension of the infinite sites model previously used for studies of codon usage bias (McVean and Charlesworth 1999). We assume a randomly mating, diploid, discrete generation population with $N$ breeding adults, and effective population size $N_e$. Over a long sequence of $m$ nucleotide sites, each site has two alternative types, $A_1$ and $A_2$, with mutation rates $u$ and $v$ from $A_1$ to $A_2$ and vice versa. $A_1$ and $A_2$ might correspond to AT versus GC base pairs, unpreferred versus preferred synonymous codons, or selectively favored versus disfavored nonsynonymous variants. If epigenetic variation is being considered, then $A_1$ and $A_2$ could be regarded as the methylated or unmethylated states of a nucleotide site or a differentially methylated region (or vice-versa). This approach, while undoubtedly oversimplified, avoids the problem of modeling mutation among all four basepairs, which is difficult to deal with except by making the unrealistic assumption of equal mutation rates in all directions (Ewens 2004, p.195).

If selection is acting, we assume semidominance, with $A_2$ having a selective advantage $s$ over $A_1$ when homozygous, although our general conclusions are probably not strongly dependent on this assumption. There is complete independence among sites (i.e. recombination is sufficiently frequent that linkage disequilibrium is negligible), and all evolutionary forces are weak, so that the standard results of diffusion approximations can be employed.

If the population is at statistical equilibrium, the mean numbers of sites in each possible state are constant over time, despite continual changes at individual sites. At any given time, some sites are fixed for the $A_1$ type, some for $A_2$, and others segregate for both. Let the equilibrium proportion of sites that are fixed for $A_1$ and $A_2$ be $f_{1f}$ and $f_{2f}$, respectively. The proportion of sites that are segregating is $f_s = 1 - f_{1f} - f_{2f}$.

### *Results for some important population parameters*

These assumptions allow the use of Wright's stationary distribution formula (Wright 1931, 1937) to describe the probability density of the frequency $q$ of $A_2$ at a site

$$\phi(q) = C \exp(\gamma q) p^{\alpha - 1} q^{\beta - 1} \tag{1}$$

where $p = 1 - q$, $\alpha = 4N_e u$, $\beta = 4N_e v$, $\gamma = 2N_e s$, and the constant $C$ is such that the integral of $\phi(q)$ between $q = 0$ and $q = 1$ is equal to 1. It is convenient to write $u$ in terms of the mutational bias parameter, $\kappa$, i.e. $u = \kappa v$, so that $\alpha = \kappa \beta$.

An explicit expression for $C$ can be obtained by noting that the integral of the other terms on the right-hand side with respect to $q$ is equal to the product of $\Gamma(\alpha) \Gamma(\beta) / \Gamma(\alpha + \beta)$ and the confluent hypergeometric function $_1F_1(a, b, z)$ (Abramowitz and Stegun 1965, p.503), with parameters $a = \beta$, $b = \alpha + \beta$, $z = \gamma$, where

$$_1F_1(a, b, z) = \sum_{i=0}^{\infty} \frac{z^i}{i!} \frac{(a)_i}{(b)_i} \qquad (2)$$

where $(x)_0 = 1$, $(x)_i = x(1+x)(2+x)...(i-1+x)$ for $i \geq 1$ (Pochhammer's symbol).

This can be seen by expanding the exponential term in equation (1) in powers of $\gamma q$, and integrating over the range 0 to 1 (Kimura *et al.* 1963).

Integrating equation (1), we have

$$C = \frac{\Gamma(\alpha + \beta)}{\Gamma(\alpha)\Gamma(\beta)} \frac{1}{_1F_1(\beta, \alpha + \beta, \gamma)} \qquad (3)$$

Furthermore, the $j$th moment of $q$ around zero, obtained from the integral of $q^j \phi(q)$ between 0 and 1, is given by

$$M_j(q) = \frac{_1F_1(\beta + j, \alpha + \beta + j, \gamma)}{_1F_1(\beta, \alpha + \beta, \gamma)} \frac{(\beta)_j}{(\alpha + \beta)_j} \qquad (4)$$

In particular, the mean frequency of $A_2$ is

$$\bar{q} = \frac{1}{(1+\kappa)} \frac{_1F_1(\beta+1, \ \alpha+\beta+1, \ \gamma)}{_1F_1(\beta, \alpha+\beta, \gamma)} \qquad (5a)$$

and the mean frequency of $A_1$ is

$$\bar{p} = \frac{\kappa}{(1+\kappa)} \frac{{}_1F_1(\beta, \alpha+\beta+1, \gamma)}{{}_1F_1(\beta, \alpha+\beta, \gamma)} \tag{5b}$$

Approximations to these expressions for the case when $\beta$ is $<< 1$ and $\kappa$ is of order 1 are derived in the Appendix. Equations (A3) imply that

$$\bar{q} = \frac{1}{[1+\kappa\exp(-\gamma)]} + O(\beta) \tag{6}$$

The left-hand side of equation (6) is equivalent to the fraction of sites that carry $A_2$ in a random sequence sampled from the population; if $A_1$ and $A_2$ correspond to unpreferred and preferred codons, respectively, this measures the frequency of preferred codons, *Fop* (McVean and Charlesworth 1999). With epigenetic variation, if $A_1$ and $A_2$ correspond to methylated and unmethylated states, $\bar{q}$ measures the fraction of unmethylated sites or regions in a random genome.

The leading term on the right-hand side of equation (6) is identical to the Li-Bulmer equation commonly used in analyses of selection on codon usage (Li 1987; Bulmer 1991). This result is, however, often derived by assuming that nearly all sites are fixed, and calculating the rate of flux between sites fixed for $A_1$ and $A_2$; $\bar{q}$ is then taken to be the frequency of sites that are fixed for $A_2$, with $1 - \bar{q}$ representing the frequency of sites fixed for $A_2$ (Bulmer 1991). This raises the question of how good an approximation we obtain by neglecting the term of order $\beta$, when the infinite sites assumption is violated, so that a significant fraction of sites are in fact segregating for $A_1$ and $A_2$.

First, we note that it is immediately obvious from (1) and equations (5) that $\overline{q}$ with $\gamma = 0$ is equal to $1/(1 + \kappa)$, so that equation (6) for this case is exact, as has long been known (Wright 1931). We can also obtain a first-order approximation to equations (A3) when $\gamma \neq 0$ by expanding in powers of $\beta$, which will be accurate when $\beta$ is sufficiently small. Neglecting second-order and higher terms in $\beta$, as will also be done in equations (8), we obtain

$$\overline{q} \approx \frac{1 - \beta \kappa g \exp(-\gamma)}{[1 + \kappa \exp(-\gamma)]} \qquad (7a)$$

where

$$g = \frac{[\gamma + \kappa \exp(-\gamma) \sum_{i=1}^{\infty} \frac{\gamma^i}{i!} a_{i+1} + \sum_{i=2}^{\infty} \frac{\gamma^i}{i!} (a_{i+1} - a_i)]}{[1 + \kappa \exp(-\gamma)]} \qquad (7b)$$

and $a_i$ is the harmonic series $1 + 1/2 + 1/3 + \ldots + 1/(i-1)$, with $i \geq 2$.

Since $a_{i+1} < i$ for $i > 1$, the first summation in the numerator of $g$ is less than the sum of $\gamma^i (i-1)!$, so that the sum is $< \gamma \exp(\gamma)$. Similarly, $a_{i+1} - a_i = 1/i$, so that the second summation is $< \exp(\gamma) - (1 + \gamma)$. It follows that $g$ is positive and $< \kappa \gamma + \exp(\gamma) - 1$. This is multiplied by $\exp(-\gamma)$ in the numerator of equation (5a), to obtain the multiplicand of $\beta \kappa$, yielding $\kappa \gamma \exp(-\gamma) + 1 - \exp(-\gamma) < \kappa + 1 - \exp(-\gamma)$. The contribution of $-\beta \kappa g \exp(-\gamma)$ to the numerator of equation (7a) is thus negative and smaller in magnitude than $\beta \kappa (1 + \kappa)$, so that the leading term in equation (6) should provide a good approximation when $\beta \kappa$ is around 0.1 or less.

For examining what happens when $\gamma$ becomes very large, it is useful to note that the Taylor's series expansion of equation (5b) for small $\beta$ yields the expression

$$\bar{p} \approx \frac{\kappa \exp(-\gamma)}{1 + \kappa \exp(-\gamma)} \left\{ 1 + \beta \left[ \sum_{i=1}^{\infty} \frac{\gamma^i}{i!} + \frac{\kappa \exp(-\gamma)}{1 + \kappa \exp(-\gamma)} \sum_{i=1}^{\infty} \frac{\gamma^{i+1} \ln(i)}{(i+1)!} \right] \right\} \qquad (8a)$$

For large $\gamma$, this gives

$$\bar{p} \approx \frac{\kappa \exp(-\gamma)}{1 + \kappa \exp(-\gamma)} \left[ 1 + \frac{\beta \exp(\gamma)}{\gamma} \right] \qquad (8b)$$

i.e.

$$\bar{p} \approx \frac{\beta \kappa}{\gamma} [1 + O(\gamma^{-1})] \qquad (8c)$$

The first term on the right-hand site of equation (8c) is equivalent to the asymptotic expression for $\bar{p}$ with large $\gamma$ given by Kimura *et al.* (1963). This implies that, for sufficiently large $\gamma$ compared with $\beta$, the mean frequency of the disfavored variant is equal to its equilibrium frequency under mutation-selection balance with $s >> u$ in an infinite population, where $p = 2u/s = 2v\kappa/s$ (Haldane 1927) , as expected intuitively. Numerical studies show that equation (8b) performs well for $\gamma > 1$ when $\beta << 1$, when it can give a good approximation when neither the leading term in equation (6) nor the Kimura *et al.* (1963) large $\gamma$ approximation are accurate (results not shown). Equation (8b) implies that the leading term in equation (6) is accurate when $\gamma << -\ln(\beta)$.

Second, the approximate frequencies of sites that are fixed for $A_1$ and $A_2$, $f_{1f}$ and $f_{2f}$, can be found from the integrals of $\phi(q)$ between 0 and $1/(2N)$ and $1 - 1/(2N)$ and 1, respectively (Ewens 2004, p.178). For large $N$, such that $\gamma N^{-1} \ll 1$, when $q$ is close to zero we have $\phi(q) = C[q^{\beta-1} + O(\gamma N^{-1}) + O(\beta N^{-1})]$, and so

$$f_{1f} \approx \int_0^{1/(2N)} \phi(q)\, \mathrm{d}q = C[\beta^{-1}(2N)^{-\beta} + O(\gamma N^{-1}) + O(\beta N^{-1})] \tag{9a}$$

$$f_{2f} \approx \int_{1-1/(2N)}^1 \phi(q)\, \mathrm{d}q = C\exp(\gamma)[(\beta\kappa)^{-1}(2N)^{-\beta\kappa} + O(\gamma N^{-1}) + O(\beta N^{-1})] \tag{9b}$$

where the terms in $O(\gamma N^{-1})$ and $O(\beta N^{-1})$ can be neglected when $N$ is sufficiently large (c.f. (Kimura 1981)). Approximations for these expressions for small $\beta$ can readily be obtained (see Appendix).

Third, the expected pairwise nucleotide site diversity, $\pi$, can be obtained from the expectation of $2pq = 2\mathrm{E}\{q - q^2\}$ between 0 and 1. From equation (4), we have

$$\mathrm{E}\{q^2\} = \frac{\beta(\beta+1)}{(\alpha+\beta)(\alpha+\beta+1)} \frac{{}_1F_1(\beta+2,\ \alpha+\beta+2,\ \gamma)}{{}_1F_1(\beta,\ \alpha+\beta,\ \gamma)} \tag{11a}$$

The expectation of $\mathrm{E}\{q - q^2\}$ is given by subtracting equation (11a) from equation (5b). Using equation (2) and simplifying, we obtain

$$\mathrm{E}\{pq\} = \frac{\dfrac{\alpha}{(\alpha+\beta+1)} + \sum_{i=1}^{\infty} \dfrac{\gamma^i}{i!}\left[\dfrac{(\beta+1)_i}{(\alpha+\beta+1)_i} - \dfrac{(\beta+1)_{i+1}}{(\alpha+\beta+1)_{i+1}}\right]}{\left[1+\kappa+\gamma+\sum_{j=2}^{\infty}\dfrac{\gamma^i(\beta+1)_{i-1}}{i!(\alpha+\beta+1)_{i-1}}\right]} \tag{11b}$$

This is equal to one-half of the expected pairwise diversity per site, $\pi$. Using the same approach as for equations (7) and (8), keeping only terms of order $\beta$ we obtain

$$\pi \approx \frac{2\beta\kappa}{\gamma} \frac{[1-\exp(-\gamma)]}{[1+\kappa\exp(-\gamma)]} \qquad (12)$$

As expected, this is identical to equation (15) of McVean and Charlesworth (1999) for the infinite sites model at statistical equilibrium, where new mutations arise only at sites that are fixed either for $A_1$ or for $A_2$. When $\gamma >> \beta$, this term converges on the deterministic value under mutation-selection balance, $2\beta\kappa/\gamma$, which corresponds to the diversity expected at deterministic mutation-selection balance with $p = 2u/s$ (see above).

In the case of neutrality, equation (11b) reduces to the following expression (Charlesworth and Charlesworth 2010, p.237)

$$\pi = \frac{2\beta\kappa}{(1+\kappa)[\beta(1+\kappa)+1]} \qquad (13)$$

As expected intuitively, the neutral diversity is always less than for the infinite sites model with a given value of $\beta$ and $\kappa$, where equation (12) with $\gamma = 0$ gives $\pi = 2\beta\kappa/(1+\kappa)$, because some new mutations arise at sites that are already polymorphic; $\pi$ approaches $2\kappa/(1+\kappa)^2$ for large $\beta$, which is the value for an infinite population at equilibrium under reversible mutation between $A_1$ and $A_2$.

### *Rate of substitution along a lineage*

The rate of substitution of new mutations along a lineage can be modeled as follows. Conditioning on a frequency $q$ of the $A_2$ variant at a site in a given generation, there is an expected number of $2Nv\kappa q$ mutations per site from $A_2$ to $A_1$, and $2Nvp$ from $A_1$ to $A_2$. The corresponding probability that the deleterious variant $A_1$ eventually becomes fixed, conditional on $p$, is $Q_1(p) = [\exp(\gamma p) - 1]/[\exp(\gamma) - 1]$ (Kimura 1962). Conditioning on this event, the probability that a new $A_1$ mutation has been fixed is $1/(2Np)$. The expected number of new $A_1$ mutations that become fixed is thus equal to $v\kappa\, p^{-1}q\, Q_1(p)$. Similarly, the conditional probability that $A_2$ eventually becomes fixed is $Q_2(q) = [1 - \exp(-\gamma q)]/[1 - \exp(-\gamma)]$; the net expected number of new $A_2$ mutations that become fixed is $v\, pq^{-1}\, Q_2(q)$. (At first sight, it would seem that this procedure cannot be applied to mutations arising in the fixed classes, and that these should be treated separately, but the argument given in the Appendix shows that it provides an accurate approximation for the situation as well.)

Integrating over all values of $q$, the net rate at which new mutations enter the population and become fixed is thus

$$\lambda = v\int_0^1 [\kappa p^{-1}qQ_1(p) + pq^{-1}Q_2(q)]\phi(q)\,\mathrm{d}q \tag{14}$$

The terms involving functions of $p$ and $q$ in the integrand are

$$p^{-1}qQ_1(p)\phi(q) = \frac{C[\exp(\gamma p) - 1]}{[\exp(\gamma) - 1]}\exp(\gamma q)p^{\alpha - 2}q^{\beta} \qquad (15a)$$

$$pq^{-1}Q_2(q)\phi(q) = \frac{C[1 - \exp(-\gamma q)]}{[1 - \exp(-\gamma)]}\exp(\gamma q)p^{\alpha}q^{\beta - 2} \qquad (15b)$$

The corresponding integrals are

$$I_1 = \left\{ \frac{\exp(\gamma) - {_1F_1}(\beta + 1, \alpha + \beta, \gamma)}{{_1F_1}(\beta, \alpha + \beta, \gamma)} \right\} \frac{\beta}{(\alpha - 1)[\exp(\gamma) - 1]} \qquad (15c)$$

and

$$I_2 = \left\{ \frac{{_1F_1}(\beta - 1, \alpha + \beta, \gamma) - 1}{{_1F_1}(\beta, \alpha + \beta, \gamma)} \right\} \frac{\alpha}{(\beta - 1)[1 - \exp(-\gamma)]} \qquad (15d)$$

Note that $_1F_1(\beta - 1, \alpha + \beta, \gamma) - 1$ has a factor of $\beta - 1$, so that the term in $\beta - 1$ in the denominator of equation (15d) cancels. At first sight, equation (15c) appears to have a singularity at $\alpha = 1$. However, by using the relation $_1F_1(a, b, z) = {_1F_1}(b - a, b, z)\exp(z)$, we find that $_1F_1(\beta + 1, \alpha + \beta, \gamma) = {_1F_1}(\alpha - 1, \alpha + \beta, \gamma)\exp(\gamma)$, so that the numerator of equation (15c) contains a factor of $\alpha - 1$, which cancels the term in the denominator.

The net rate of substitution is given by

$$\lambda = v(\kappa I_1 + I_2) \qquad (16)$$

As $\gamma$ approaches zero, equations (15) and (16) imply that $\lambda$ tends to $2\nu\kappa/(1 + \kappa)$; this is independent of the population size and is identical to the infinite sites expression with neutrality at statistical equilibrium under reverse mutation (Charlesworth and Charlesworth 2010, p.274), as expected from the fact that the equilibrium neutral substitution rate is equal to the net mutation rate for any class of mutational model (Kimura 1968).

When $\alpha$ and $\beta$ are sufficiently small, the main contributions to $\lambda$ come from the two fixed classes, so that the initial frequencies of the new mutations can be equated to $1/(2N)$, when $o(\beta)$ terms in $I_1$ and $I_2$ are neglected. Using the above result that the frequencies of the fixed classes are equal to the infinite sites values multiplied by a factor $1 - O(\beta)$, the infinite sites expression is for the case of selection is recovered, neglecting higher order terms in $\beta$ (Equation 6.11 of (Charlesworth and Charlesworth 2010, p.275) ). Again, this implies that, as expected, the infinite sites model provides a good approximation for the rate of substitution with sufficiently small $\beta$.

There are two different ways in which we can determine the ratio of the value of $\lambda$ with $\gamma > 0$ to that for a neutral standard, thereby removing the dependence on the mutation rate term in equation (16). First, $\lambda$ with selection can be compared to its value at statistical equilibrium with the same value of $\alpha$ and $\beta$. This would be appropriate for comparing rates of evolution at putatively neutral sites in a given genomic region with sites that are potentially under purifying selection, without making any corrections for differences in base composition; this is often done when comparing nonsynonymous and synonymous rates of substitution across different genes by statistics such as $K_A/K_S$. Second, $\lambda$ with selection can be compared with the neutral rate conditioned on the same mean frequencies of $A_1$ and $A_2$ along the sequence as for the selected sites; this corresponds to methods that compare probabilities of substitution between the same pairs of nucleotides in contexts when these are putatively selected versus putatively neutral (Halligan *et al.* 2004; Eory *et al.* 2010).

### Numerical results for the population parameters

Numerical results generated from the above formulae are presented in Figures 1 and 2. Figure 1 illustrates the dependence of the following variables on the scaled mutation rate ($\beta$) and the scaled intensity of selection ($\gamma$), assuming a mutational bias ($\kappa$) of 2 towards the deleterious variant at a site): the mean frequency per site of the deleterious variant $A_1$ ($\bar{p}$), the expected diversity ($\pi$), the expected proportion of sites that segregate for variants ($f_s$), and the above two measures of the rate of substitution relative to neutral expectation. Figure 2 illustrates the dependence of $\bar{p}$ and $\pi$ on $\beta$ at a finer scale, for different values of $\kappa$ and $\gamma$. For clarity, the infinite sites values for $\bar{p}$ and the relative rates of substitution are not shown; with selection, the infinite sites values for these parameters are close to their values when $\beta = 0.002$. With neutrality, the exact value of $\bar{p}$ is always equal to the infinite sites value, and is independent of $\beta$ for a given value of $\kappa$. With selection and low $\beta$ (0.002 or 0.02), it can be seen that agreement with the infinite sites predictions is quite good for both these values despite the fact that the proportion of sites that are segregating can be quite substantial with $\beta = 0.02$; the second-order approximation of equations (7) gives very close agreement even for $\beta = 0.2$ with weak selection, but diverges for $\beta > 0.2$ when $\gamma > 0.5$ (results not shown), whereas the value of $\bar{p}$ departs quite seriously from the infinite sites values at $\beta = 0.2$ when $\gamma > 5$. A similar pattern of departure from the infinite sites value holds for $\pi$, except when selection is strong ($\gamma = 5$ or $50$), when agreement is still good at $\beta = 2$; this is because the exact diversity and the infinite sites value both approach the deterministic value under mutation-selection balance when $1 << \gamma$ and $\beta << \gamma$ (see equation 12). Somewhat surprisingly, the infinite sites and exact values of the proportion of sites that are segregating always agree well.

Perhaps the most interesting result to emerge is that, with $\kappa > 1$, the rate of substitution relative to neutral expectation can exceed one when there is moderate selection and mutational bias towards the deleterious variants. This has long been known to apply to the infinite sites model when the "uncorrected" relative rate is used and when there is mutational bias (Eyre-Walker 1992; McVean and Charlesworth 1999), which can cause serious problems for phylogenetic inferences concerning selective constraints (Lawrie *et al.* 2011). As shown in the Appendix, the "corrected" relative rate is always expected to be less than one under the infinite sites assumption (see equation A8). But with sufficiently high $\beta$, the corrected relative rate can exceed one, even for $\gamma = 5$, and can be only just below one for lesser values of $\beta$. The reason for this seemingly paradoxical result is presumably the fact that nearly all sites are segregating when $\beta$ is high; when $\bar{p}$ is sufficiently high because mutation and drift are overcoming selection, there is a substantial chance that a new mutation to the favorable variant $A_2$ can arise at a segregating site, which has a higher chance of fixation than a neutral variant and hence contributes to an elevated substitution rate. With sufficiently strong mutational bias, $\bar{p}$ can be much greater than ½, so that the contribution from the enhanced fixation probability of favorable mutations outweighs the lower contribution from the fixation of deleterious mutations.

As was previously shown by McVean and Charlesworth (1999) for the infinite sites model, the equilibrium diversity with selection can also considerably exceed the neutral equilibrium value with the same mutational parameters, when there is a mutational bias towards deleterious alleles (see also Kondrashov *et al.* (2006)). For example, in Figure 1, with $\gamma = 5$ and $\beta = 2$, $\pi = 0.43$ but is 0.38 for the neutral case; with $\beta = 20$, and $\gamma = 50$ the values are 0.49 and 0.44, respectively. In this case, there is no meaningful way of correcting for differences in base composition between the neutral and selected sites when there are substantial departures from the infinite sites assumption, since the diversity in the neutral case is not related to the mean allele frequency in a simple way.

### *Properties of a sample from a population*

This raises the question of the extent to which the properties of a sample of alleles from a population are affected by deviations from the infinite sites assumption. With the above model, the probability that a sample of $n$ alleles segregates for $k$ $A_2$ variants at a site and $n - k$ copies of $A_1$ can be obtained from the corresponding binomial distribution with parameter $q$, integrated over $\phi(q)$, and takes the form

$$p(k) = \binom{n}{k} C \int_0^1 \exp(\gamma q)\,(1-q)^{\alpha+n-k-1}\, q^{\beta+k-1} \mathrm{d}q \qquad (17a)$$

where *C* is given by equation (3) (McVean and Charlesworth 1999; Desai and Plotkin 2008).

Using the properties of the confluent hypergeometric function, this yields

$$p(k) = \binom{n}{k} \frac{{}_1F_1(\beta+k, \alpha+\beta+n, \gamma)\,(\beta)_k\,(\alpha)_{n-k}}{{}_1F_1(\beta, \alpha+\beta, \gamma)\,(\alpha+\beta)_n} \quad (0 < k < n) \qquad (17b)$$

$$p(0) = \frac{{}_1F_1(\beta, \alpha+\beta+n, \gamma)\,(\alpha)_n}{{}_1F_1(\beta, \alpha+\beta, \gamma)\,(\alpha+\beta)_n} \qquad (17c)$$

$$p(n) = \frac{{}_1F_1(\beta+n, \alpha+\beta+n, \gamma)\,(\beta)_n}{{}_1F_1(\beta, \alpha+\beta, \gamma)\,(\alpha+\beta)_n} \qquad (17d)$$

The proportion of sites that are observed to be segregating is

$$p_{seg} = 1 - p(0) - p(n) \qquad (17e)$$

The conditional site frequency spectrum (SFS for segregating sites can be obtained by dividing equation (17b) by (17e). The folded SFS for segregating sites (which describes the numbers of variants of either type at frequencies 1 up to $0.5n + 1$ (*n* odd) or $0.5n$ (*n* even) can also readily be obtained.

Equations (17) can readily be used to obtain the theoretical values of standard sample statistics, such as the diversity per site ($\pi$) (Tajima 1983), Watterson's $\theta_w = p_{seg}/a_n$ (Watterson 1975) and Tajima's *D* (Tajima 1989b), using the standard formulae for these quantities. A well-known problem with Tajima's $D$ is the fact that its magnitude is strongly dependent on both the level of variability in the population and on the length of sequence used to estimate it (Tajima 1989b). Langley *et al.* (2014) proposed the use of the summary statistic $\Delta_\pi = (\pi - \theta_w)/\theta_w$ for measuring the extent of departure of the SFS from the infinite sites neutral equilibrium expectation, which should not suffer from these problems. Another summary statistic for this purpose is provided by the proportion of singleton variants among segregating sites, given by

$$p_{sn} = [p(1) + p(n-1)]/p_{seg} \qquad (17f)$$

(This is closely related to the widely used $D$ statistic of Fu and Li (1993).)

Use of the series expression for the confluent hypergeometric function allows rapid computation of all relevant statistics; to avoid overflow when $\gamma$ is large, however, it is necessary to use logarithms of the individual terms and partial sums of the series (this requires the selection model to be defined such that $\gamma > 0$). A FORTRAN program is available on request to BC.

Table 1 displays some examples of such computations, for the case of a mutational bias of 2 towards deleterious mutations, for a subset of the parameter values used in Figure 1. The expected $\pi$ values are not shown, since these are the same as the population diversities given in Figure 1. Figure 3 show the folded SFSs for some chosen examples, using a sample size of 20 alleles. It can be seen that a high $\beta$ value (20) means that the proportion of sites that are found to be segregating ($p_{seg}$) is effectively 100%, even for $\gamma$ as high as 50 and a sample size ($n$) of 20. A moderate $\beta$ value (0.2) behaves similarly in the neutral case with a sample size of 200, but otherwise is associated with a $p_{seg}$ of less than 80% (and is as low as 13% for $n = 20$ and $\gamma = 50$). With neutrality or weak selection ($\gamma \leq 5$), moderate or high values of $\beta$ cause a distortion of the SFS towards a much lower proportion of singletons ($p_{sn}$) and higher Tajima's $D$ and $\Delta_\pi$ than is expected with the infinite sites model. Even for $\gamma = 50$, a very low $p_{sn}$ and a positive $D$ are found when $\beta = 20$. This reflects the tendency of high $\beta$ values to push the distribution of $q$ towards intermediate frequencies, which has long been known (Wright 1931). Some analytical approximations for $p_{seg}$ are derived in File S1.

## Discussion

The results described above have some important implications for the interpretation of data on DNA sequence variation and evolution when there is "hyperdiversity", i.e., the scaled mutational parameter ($\beta$ in the notation used here) is sufficiently large that the infinite sites model does not accurately describe patterns of variation within populations. Recent surveys of DNA sequence polymorphisms show that that such hyperdiversity is more common than previously thought, even in multicellular organisms (Cutter *et al.* 2013). In addition, given the evidence from studies of organisms like *Arabidopsis thaliana* and maize that epigenetic variants such as methylated cytosines can be transmitted fairly stably through meiosis, but have origination and disappearance rates that are several orders of magnitude higher than those of nucleotide variants (Johannes *et al.* 2009; Becker *et al.* 2011; Schmitz *et al.* 2011; Lauria *et al.* 2014), the patterns described above are relevant to population level studies of some classes of epigenetic variants.

### Distortion of the SFS with hyperdiversity

As was pointed out about twenty years ago in the context of human mitochondrial DNA sequence variability (Bertorelle and Slatkin 1995; Aris-Brosou and Excoffier 1996; Tajima 1996; Yang 1996), a major effect of a high scaled mutation rate ($\beta$ in the notation used here) is that more intermediate frequency variants will be present at polymorphic sites in a sample from a population than under the equilibrium infinite sites model. In particular, for a stationary population at equilibrium between drift and the input of neutral or nearly neutral mutations, the expected values of Tajima's $D$ statistic ($D_T$) and the $\Delta_\pi$ statistic proposed by Langley *et al.* (2014) are positive rather than slightly negative or zero, respectively, as expected under the infinite sites model (Tajima 1989) – see Figure 1 and Table 1). This reflects the fact that the expected value of the pairwise diversity per site ($\pi$) is greater than the expected value of the measure of diversity based on the number of segregating sites at a locus ($\theta_w$). As can be seen from Table 1, this effect is quite noticeable even for $\beta$ as low as 0.02 when selection is absent or weak, and small positive values of $D_T$ and $\Delta_\pi$ are found with neutrality even when $\beta = 0.002$ (of the order of 1% with $n = 200$).

   With very high values of $\beta$, a positive Tajima's $D$ can occur even with quite strong purifying selection (a scaled selection parameter $\gamma$ of 50) can be associated with (Table 1). A site frequency frequency spectrum (SFS) with an excess of intermediate frequency variants at loci across the genome is usually interpreted as indicating a recent population bottleneck or a subdivided population, e.g. Staedler *et al.* (2009). False positive results for tests for bottlenecks and/or subdivision may thus be obtained if infinite sites rather than finite sites models are applied to hyperdiverse populations or epigenetic variation, even when moderately strong purifying selection is acting. Given that very small positive mean values across sites of statistics such $\Delta_\pi$ can be statistically significant with genomic scale data and large sample sizes, caution should be exercized in using infinite sites predictions for such datasets. The suggested criterion for hyperdiversity of $\pi$ or $\theta_w$ of 5% for using finite sites models rather than the infinite sites model (Cutter *et al.* 2013) may be too high for such data.

This raises the question of whether there is indeed evidence for the expected pattern of a skew of the SFS spectrum towards intermediate frequency variants. In the study of *Caenorhabditis* sp.5 by Cutter *et al.* (2012), where the within-population diversity at synonymous sites is about 0.08, Tajimas' *D* values for "scattered" samples (where one allele per locus was sampled from each of 13 locations, in order to minimize departure from the standard coalescent process (Wakeley 2000)) were nearly all positive, with a mean of 0.28. This is consistent with the coalescent simulations of Cutter *et al.* (2012), who used the SIMCOAL2 program of Laval and Excoffier (2004) with a finite-sites model with equal mutation rates among all four possible nucleotides (A. Cutter and L. Excoffier, pers. comm.). The model used here gives an expected value of Tajima's *D* of approximately 0.10 with $\gamma = 0$ or 0.5 and a mutational bias of 2, assuming a sample size of 13 and 150bp per locus (corresponding approximately to the numbers of synonymous sites in the study). At least qualitatively, this species thus fits the expectation under hyperdiversity for DNA sequence variability.

In contrast, the synonymous SFS in the much more hyperdiverse species *C. brenneri* is biased towards low frequency variants, with a mean Tajima's *D* of –0.56 over 23 loci with a average of approximately 150bp per locus (Dey *et al.* 2103, Table S3), again using scattered sampling. Similarly, in the only detailed survey of epigenomic variation published to date, that of approximately 200 northern European accessions of *Arabidopsis thaliana* (Schmitz *et al.* 2013, Supplementary Table 9), the SFS for single methylated versus nonmethylated cystosines is also highly skewed towards low frequency variants. The lack of linkage disequilibrium between this class of variants and SNPs suggests that these epigenetic variants are not caused by nucleotide site variants associated with methylation status, but represent true heritable epialellic variation (Schmitz *et al.* 2013).

There are several possible reasons for this sharp disagreement between the theoretical predictions and these observations. One is that demographic effects, such as a recent population expansion, mean that predictions based on the assumption of a stationary population are overwhelmed by the well-known excess of rare variants associated with expansion (Tajima 1989a; Slatkin and Hudson 1991). This is ruled out for the case of epigenetic variation in *A. thaliana,* because the SFS for SNPs is far less biased towards rare variants (Schmitz *et al.* 2013), but remains possible for *C. brenneri*. The second possibility is that purifying selection is sufficiently strong to  skew the SFS towards rare variants. This seems unlikely in the case of *C. brenneri*, where the estimates of the overall $\gamma$ for synonymous sites suggests a value close to 0.5 (Dey *et al.* 2103), which is insufficient to cause a skew towards rare variants (see Table 1). This explanation is more plausible for the *A. thaliana* example, since high levels of methylation of cytosines are non-randomly distributed across the genome, and are especially prevalent in transposable element sequences where methylation is important for their silencing (Schmitz *et al.* 2013). It is therefore very likely that the methylated states in such sequences are favored by selection. Another possibility is that methylation is selectively neutral, and the differences between genomic regions simply reflect different levels of mutational bias, either towards or against methylation. Calculations using the biallelic model show that extreme mutational bias at neutral or nearly neutral sites can overcome the skew of the SFS towards intermediate frequency variants (results not shown). The published results of mutation accumulation experiments in *A. thaliana* (Becker *et al.* 2011; Schmitz *et al.* 2011) do not shed much light on the question of the extent of the direction and magnitude of mutational bias, since the experimental design ascertains sites for which at least one of the mutation accumulation lines contains a methylated cytosine at the site in question. It is thus strongly biased towards detecting variants at which the original state was methylation, making it hard to determine the rate of mutation towards methylation. Distinguishing between these possible interpretations is a challenging task, and will require the use of numerical models that incorporate past population size changes and population structure.

### *Limitations of the biallelic model*

It is important to note that the biallelic model used here, which is similar to that used by Bertorelle and Slatkin (1995) and Desai and Plotkin (2008), is likely to underestimate the effect of hyperdiversity on the SFS, since the presence of more than two variants at a segregating site will result in higher $\pi$ but not $\theta_w$. On the other hand, the infinite alleles assumption apparently used by Aris-Brosou and Excoffier (1996) means that the upper limit to $\pi$ is 1, whereas in reality there is a maximum of four segregating variants per site, leading to an upper limit to $\pi$ of 3/4 (when all four variants are present at equal frequencies), as opposed to 1/2 for the biallelic model used here. Given the almost universal existence of mutational biases towards transitions versus transversions, and for GC to AT versus AT to GC mutations, the upper limit is in practice likely to be considerably smaller than 1, so that the biallelic model with modest mutational bias probably provides a reasonably good guide to the values of measures of skew in the SFS.

An intermediate situation is provided by assuming a *K*-allele model (Ewens 2004 pp.192-200) with $K = 4$, corresponding to equal mutation rates among all 4 nucleotide states at site (Tajima 1996; Yang 1996; Desai and Plotkin 2008). Under neutrality the exchangeability of the different nucleotides under this model means that the probability density $\phi(q_i)$ for the frequency $q_i$ of a variant of type $i$ ($i = 1 - 4$) is proportional to $(1 - q_i)^{\theta - 1} q_i^{(\theta/3) - 1}$, where $\theta$ is the net mutation rate per site, i.e. $\phi(q_i)$ follows a beta distribution with parameters $\theta$ and $\theta/3$ (Tajima 1996). With semi-dominant selection with type $i$ having a selective advantage $s$ over all other variants, which are assumed to be selectively equivalent to each other, this expression is simply multiplied by $\exp(\gamma q_i)$.

Following Tajima (1996), these assumptions allow simple analytical formulae for the sample statistics used above to be obtained for the case of neutrality:

$\pi = \theta / [1 + (4\theta / 3)]$, $p_{seg} = 1 - [S_{n-1}(\theta / 3) / S_{n-1}(4\theta / 3)]$, and

$p_{sn} = n\theta S_{n-2}(\theta) / P_{seg} S_{n-1}(4\theta / 3)$, where $S_k(x) = (1 + x)(2 + x)...(k + x)$. These can be compared with the statistics obtained from the biallelic model in Table 1, setting $\theta$ to the equilibrium infinite sites neutral diversity with reverse mutation $2\beta\kappa / (1 + \kappa) = 4\beta/3$ (with $\kappa = 2$) to obtain comparable net scaled mutation rates per site. As expected, for very low $\theta$, the two models yield similar results, but even with $\beta = 0.02$ the 4-allele model gives noticeably higher expected values of Tajima's $D$ and $\Delta_\pi$ ; e.g. with a sample size of 20 and $\beta = 0.02$, the values of Tajima's $D$ and $\Delta_\pi$ are 0.069 and 0.022, respectively, versus 0.038 and 0.016 for the biallelic model. With a sample size of 20 and $\beta = 0.2$, the values of Tajima's $D$ and $\Delta_\pi$ for the 4-allele model are 0.61 and 0.20, respectively, compared with 0.32 and 0.11 for the biallelic model; values of $D$ and $\Delta_\pi$ much greater than twice the biallelic values can be generated by the 4-allele model when $\beta$ is large, reaching 4.7 and 1.6, respectively, with $\beta = 20$. The proportion of singletons behaves rather differently under the 4-allele model; it can even increase with $\beta$ up to some upper limit, after which it declines, and is always higher than for the biallelic model (e.g. 0.36 versus 0.22, respectively, for $\beta = 0.2$ and $n = 20$; 0.17 versus 0.01 for $\beta = 20$). This behavior presumably reflects the fact that there are four possible variants at each site that can behave as singletons in the case of the 4-allele model, and the above formula simply sums over the probabilities that each one of these is a singleton, regardless of the status of the other three possible variants at the same site. A statistic such as $\Delta_\pi$ is thus probably a better summary of the skew of the SFS than the proportion of singletons when a substantial fraction of polymorphic sites segregate for more than two variants, unless variants are collapsed into biallelic alternatives such as GC versus AT basepairs.

For studying situations with multiple alleles per nucleotide and non-equilibrium demography, numerical methods such as that of Zeng (2010) will be needed.

### *Some other implications*

One difficulty with interpreting the results of population surveys of epiallelic variation is that it is impossible to know whether sites that lack epigenetic marks in all individuals sampled are potentially capable of acquiring them. This means that the denominator in per-site statistics such as $\pi$ and $\theta_w$ is unknown, making it hard to apply standard population genetics methods to this kind of data. Fortunately, however, with high β values (> 0.2), nearly all sites capable of mutation will be found to be segregating in a large sample, even with a scaled selection strength as high as $\gamma = 5$ (Table 2); thus, the majority of sites capable of epimutations can be identified from population surveys, unless strong purifying selection is acting. Population surveys could, therefore, be a valuable tool for the characterization of the epigenome.

Another finding that is relevant for both hyperdiverse DNA sequence variation and hypermutable epigenetic variation is the fact that substitution rates for sites under purifying selection may be close to, or even greater than rates at neutral sites with high β values. As described above, this may occur even after corrrecting for the effects of differences in base composition between neutral and selected sites (Figs 1 and 2). This lack of sensitivity of substitution rates to the strength of purifying selection is consistent with the patterns described by Cutter *et al.* (2013), where there is only a weak relation between codon usage bias and a measure of synonymous site divergence in the hyperdiverse species *Ciona savigni*. Similarly, diversity at sites subject to weak purifying selection is expected to show a non-linear pattern of relationship with $\gamma$, such that $\pi$ increases with $\gamma$ when sites are close to neutral, and then declines again with as $\gamma$ approaches or exceeds 1; the range of $\gamma$ values over which there is an increase is broader for large β (Figure 2). Synonymous diversity of genes in *C. brenneri* does indeed show a quadratic relation with the frequency of optimal codons, such that genes with approximately 50% optimal codons have the highest diversity values (Asher Cutter, personal communication).

With $\gamma$ values typical of those reported from studies of selection on codon usage ($\gamma$ =

1 or less), the standard Li-Bulmer equation (Li 1987; Bulmer 1991) tends to overestimate the expected level of codon bias, as measured by the mean frequency of the favored allelic type ($\bar{q}$), when β > 0.02. For example, with $\gamma$ = 1 and $\beta$ = 0.2, the exact value of $\bar{q}$ from equation (5a) is 0.49 compared with the Li-Bulmer infinite sites prediction of 0.58, while the second-order approximation from equations (7) gives 0.44.

Analyses of codon usage in hyperdiverse species that use codon usage data to estimate

$\gamma$ , (see Sharp *et al.* (2010)), should probably use the exact expression. It is interesting in

this context to note that there is there is only a small difference in the mean level of codon usage bias between *C. brenneri* and *C. remanei*, despite an approximately three-fold difference in synonymous site diversity (Asher Cutter, personal communication).

This raises the question of whether the purifying selection model used here is appropriate for codon usage, or whether a model of stabilizing selection (Kimura 1981) is more realistic, since the latter means that $\gamma$ is insensitive to $N_e$ over a wide range of parameter values, provided that there is mutational bias (Charlesworth 2013). The behavior of this model with hyperdiversity is, therefore, worth studying.

***Sample versus population distributions; reversible versus irreversible mutational models***

Lawrie *et al.* (2013) have proposed that, for the irreversible mutation model with large $n$ (see below), it is computationally more efficient to replace the sampling formula for $p(k)$ by multiplying the probability density $\phi(q)$ by $1/n$, i.e. d$q$ is equated to $1/n$, to obtain the probability of obtaining an allele frequency $q = k/n$ in the sample. The corresponding procedure for the reversible mutation model yields

$$p_{pop}(k = qn) = C \exp(\gamma q)(1-q)^{\alpha-1} q^{\beta-1} n^{-1} \qquad (18)$$

This relation can be used to obtain statistics analogous to those described by equations (17b) to (17f), using summation over all values of $k$ between 1 and $n-1$ to obtain statistics such as the proportion of segregating sites, the expectations of the diversity measures $\pi$ and $\theta_w$, and the proportion of singletons.

Table 2 displays the results of computations using this formula for $n = 200$. These show that the use of the population distribution instead of the exact sampling distribution overestimates the above statistics, unless selection is extremely strong ($\gamma \gg 50$). Tajima's $D$, on the other hand, is usually underestimated, reflecting the overestimation of the proportion of singletons. These results can be compared with the approach of Lawrie *et al.* (2013), by noting that, for strong selection and small $\beta$, the reversible mutation model converges on the irreversible mutation model, which assumes that mutations are all from $A_2$ to $A_1$, so that the relevant scaled mutation rate is $\alpha$ (the convergence can never be exact, since the irreversible mutation model cannot achieve true stationarity).

The results for $\gamma = 50$ and 500 in Table 2 should thus correspond closely to those for the irreversible mutation model, especially for $\beta = 0.002$.

This was confirmed by direct calculation of the properties of the irreversible mutation model. The sample SFS was determined using the series expansion derived by Welch *et al.* (2008, equation 8), which is analogous to equations (17), replacing their $\theta$ with $\alpha$. A similar series expansion can be obtained for the SFS and proportion of segregating sites generated from the population distribution (see Appendix). Comparisons of the SFSs for segregating sites for the reversible and irreversible mutation models shows close agreement between the two when $\beta = 0.002$, even for $\gamma$ as low as 5; with $\beta \geq$ 0.02, agreement is less close. For example, with $\gamma = 5$ the proportions of singletons using the sample SFS are 0.244 and 0.226 for the reversible model with $\beta = 0.002$ and 0.02, respectively, compared with 0.248 for the irreversible model. The population distributions give very similar values. The proportions of segregating sites from the sample distribution for the reversible model with $\gamma = 5$ are 0.016 and 0.146 for $\beta = 0.002$ and 0.02, respectively, compared with 0.016 and 0.158 for the irreversible model; the corresponding population values for the reversible model are 0.029 and 0.201 for $\beta =$ 0.002 and 0.02, respectively, while the corresponding values for the irreversible model are 0.032 and 0.315. In general, for $\beta = 0.02$, the irreversible model tends to give a larger discrepancy between the sample and population distribution values of the proportion of segregating sites than that seen under the reversible model. It seems, therefore, that methods such as those of Lawrie *et al.* (2013), which make use of the proportion of segregating sites to estimate the intensity of selection, will be substantially biased by using the population distribution rather than the sample distribution, especially with the irreversible model, even when the SFSs for the two distributions are very similar. Indeed, since computationally efficient algorithms for generating the SFS for samples in the case of a stationary population are readily implemented, it is unnecessary to use the short-cut proposed by Lawrie *et al.* (2013).

**Acknowledgments**

We thank Deborah Charlesworth, Asher Cutter and Kai Zeng for their comments on the manuscript, and Asher Cutter and Laurent Excoffier for valuable discussions.

**Appendix**

**The expected number of new mutations that arise at a segregating site**

Assume that we have a nucleotide site that is segregating for a neutral mutation that arose at an initial frequency of $1/(2N)$. Let the probability that this variant mutates to an alternative nucleotide be $u$ per generation (this includes the possibility that it reverts to the ancestral state); let the probability that the ancestral variant mutates to another state be $v$ (this includes the possibility that the mutation is identical in state to the variant that is already segregating). If the frequency of the mutation in the population in a given generation is $x$, the expected total number of mutational events is $2N[ux + v(1 - x)]$. The expected time that the original mutation spends in the frequency interval $x$ to $x + \mathrm{d}x$ is given approximately by $4N_e/(1 - x)$ for $0 < x \leq 1/(2N)$, and $2N_e/(Nx)$ for $1/(2N) < x \leq 1$ (Ewens 2004, p.160). The total expected number of new mutations that arise during the sojourn of the mutation in the population is thus

$$4NN_e \left\{ \int_0^{1/(2N)} \frac{2[ux + v(1 - x)]}{(1 - x)} \, \mathrm{d}x + \int_{1/(2N)}^1 \frac{[ux + v(1 - x)]}{Nx} \, \mathrm{d}x \right\} \approx 4N_e[u + v\ln(2N)]$$

(A1)

**Approximations to equations (5) with small $\alpha$ and $\beta$**

Equation (5a) is equivalent to

$$\bar{q} = \frac{\left[ 1 + \sum_{i=1}^{\infty} \frac{\gamma^i \, (\beta + 1)_i}{i! \, (\alpha + \beta + 1)_i} \right]}{\left[ 1 + \kappa + \gamma + \sum_{i=2}^{\infty} \frac{\gamma^i (\beta + 1)_{i+1}}{i! (\alpha + \beta + 1)_{i+1}} \right]}$$

(A2)

We can write terms of the form $(\beta + i - j)/(\alpha + \beta + i - j)$ as $1 - [\beta\kappa/(i - j)] + O(\beta^2)$;

keeping only $O(\beta)$ terms, we have

$$\bar{q} \approx \frac{\{1 + \sum_{i=1}^{\infty} \frac{\gamma^i}{i!} \prod_{j=1}^{i} [1 - \frac{\beta\kappa}{(i+1-j)}]\}}{\{1 + \kappa + \gamma + \sum_{i=2}^{\infty} \frac{\gamma^i}{i!} \prod_{j=1}^{i-1} [1 - \frac{\beta\kappa}{(i-j)}]\}} \tag{A3a}$$

or

$$\bar{q} \approx \frac{\{1 + \sum_{i=1}^{\infty} \frac{\gamma^i}{i!} \exp(-\beta\kappa a_{i+1})\}}{\{1 + \kappa + \gamma + \sum_{i=2}^{\infty} \frac{\gamma^i}{i!} \exp(-\beta\kappa a_i)\}} \tag{A3b}$$

where $a_i = 1 + 1/2 + 1/3 + \dots 1/(i - 1)$ $(i \geq 2)$. The exponential terms in the numerator

and denominator of equation (A3b) can thus be replaced by $1 + O(\beta)$, yielding equation

(6) of the text.

## Approximations for the frequencies of the fixed classes

Assuming that $\alpha \ll 1$ and $\beta \ll 1$, and employing the approximations used in equation

(A3b), we find that

$$f_{1f} \approx \frac{\Gamma(\alpha + \beta)}{\Gamma(\alpha)\Gamma(\beta)} \left\{ \frac{1 + \kappa}{[1 + \kappa + \gamma + \sum_{i=2}^{\infty} \frac{\gamma^i}{i!} \exp(-\beta\kappa a_i)]} + o(\beta) \right\} \beta^{-1}(2N)^{-\beta}[1 + O(\gamma N^{-1}) + O(\beta N^{-1})]$$

$$(A4a)$$

Similarly,

$$f_{2f} \approx \frac{\Gamma(\alpha + \beta)\exp(\gamma)}{\Gamma(\alpha)\Gamma(\beta)} \left\{ \frac{1 + \kappa}{[1 + \kappa + \gamma + \sum_{i=2}^{\infty} \frac{\gamma^i}{i!} \exp(-\beta\kappa a_i)]} + o(\beta) \right\} (\beta\kappa)^{-1} (2N)^{-\beta\kappa} [1 + O(\gamma N^{-1}) + O(\beta N^{-1})]$$

$$(A4b)$$

We can use the fact that $\Gamma(1 + x) = x\,\Gamma(x)$ to write $\Gamma(x) = \Gamma(1 + x)/x$. For small $x$, the representation of the gamma function as an infinite product (Abramowitz and Stegun 1965) implies that $\Gamma(1 + x)(1 - c\,x) + o(x)$, where $c \approx 0.577$, is Euler's constant, and similarly for the other gamma integrals. We can thus approximate $\Gamma(x)$ for small $x$ by $1/x$, and the term involving gamma functions in equations (9) is then $\kappa\beta/(1 + \kappa)[1 + O(\beta)]$. Using a similar approximation to that used in equations (7), and neglecting higher-order terms in $\beta$, we obtain

$$f_{1f} \approx \frac{\kappa \exp(-\gamma)(2N)^{-\beta}}{[1 + \kappa \exp(-\gamma)]} \left\{ 1 + \frac{\beta\kappa h \exp(-\gamma)}{[1 + \kappa \exp(-\gamma)]} \right\} [1 + O(\gamma N^{-1}) + O(\beta N^{-1})]$$

$$(A4c)$$

$$f_{2f} \approx \frac{(2N)^{-\beta\kappa}}{[1 + \kappa \exp(-\gamma)]} \left\{ 1 + \frac{\beta\kappa h \exp(-\gamma)}{[1 + \kappa \exp(-\gamma)]} \right\} [1 + O(\gamma N^{-1}) + O(\beta N^{-1})]$$

$$(A4d)$$

where

$$h = \sum_{i=2}^{\infty} \frac{\gamma^i}{i!} a_i \qquad\qquad \text{(A4e)}$$

The higher-order terms in $\beta$ vanish when $\gamma = 0$, suggesting that these expressions are good approximations when $\beta$ and $\gamma$ are both small. More rigorously, for finite $i$, $a_i$ is less than some constant $A$, which is approximately equal to $\ln(i-1)$. If terms in $i > k$ can be neglected in the sum that defines $h$, $h < \ln(k) \exp(\gamma)$, so that $h \exp(-\gamma) < \beta \kappa \ln(k)$, where $\ln(k)$ is a small multiple of one unless $\gamma$ is very large.

In addition, for arbitrary $\gamma$, the terms involving $(2N)^{-\beta}$ in equations (A4c) and (A4d) are equal to $1 - \beta \ln(2N) + O(\beta^2)$ and $1 - \beta \kappa \ln(2N) + O(\beta^2)$, respectively. Provided that $\ln(2N)$ is of order one, $f_{f1}$ and $f_{f2}$ are each equal to their respective infinite sites value, multiplied by a factor $1 - O(\beta)$, implying that the infinite sites values provide a good approximation unless $\beta >> 0$.

**Fixations of mutations**

Consider first the case of $A_2$ to $A_1$ mutations that arise at a site that was initially fixed for $A_2$. We approximate the frequency of this fixed class, $f_{2f}$, by the integral in equation (9b). The fixation probability of an $A_1$ mutation with initial frequency $1/(2N)$ when $N$ is large is $\gamma (2N)^{-1} [\exp(\gamma) - 1]^{-1} + O[\gamma^\square (2N)^{-2}]$, so that the net number of new $A_2$ mutations that arise in a given generation and are expect to become fixed is $2N\kappa v f_{2f} \{ \gamma/(2/N)[\exp(\gamma) - 1]^{-1} + O[\gamma^\square (2N)^{-2}] \} = k v f_{2f} \{ \gamma [\exp(\gamma) - 1]^{-1} + 2N\, O[\gamma^\square (2N)^{-2}] \}$. Using the same approximation for $Q_1$, and the fact that $q$ is close to one in equation (9b), the corresponding formula from equations (14) and (15a) is

$$\kappa v \int_{1-1/(2N)}^{1} Q_1(p)\, p^{-1} q\, \phi(q)\, \mathrm{d}q = \kappa v \Big\{ \gamma[\exp(\gamma)-1]^{-1} \int_{1-1/(2n)}^{1} \phi(q)\, \mathrm{d}q + O[\gamma^2 (2N)^{-2}] \Big\}$$

$$(A5)$$

Provided that $2N$ is sufficiently large in relation to $\gamma$, so that the higher order terms in $\gamma(2N)^{-1}$ can be ignored, the two results are equivalent.

The following argument can be used for the other end of the frequency range. In this case, there is no contribution from the class fixed for $A_1$ mutations (frequency $f_{1f}$, as given by equation (9a)) to the fixation of new $A_1$ mutations. The corresponding formula from equations (16) and (17a) is

$$\kappa v \int_{0}^{1/(2N)} Q_1(p)\, p^{-1} q\, \phi(q)\, \mathrm{d}q = \kappa v\, (2N)^{-1} \{1 + O[(1+\gamma)(2N)^{-1}]\} \quad (A6)$$

Again, provided that $2N$ is sufficiently large in relation to $\gamma$, the two results are equivalent.

Parallel arguments can be used for the fixation of new $A_2$ mutations.

**The relative rate of substitution under the infinite sites assumption**

At equilibrium between mutation, drift and selection, the frequencies of sites fixed for $A_1$ and $A_2$ under the infinite sites model are approximated by $\kappa \exp(-\gamma)/[1 + \kappa \exp(-\gamma)]$ and $1/)/[1 + \kappa \exp(-\gamma)]$, respectively (Li 1987; Bulmer 1991; McVean and Charlesworth 1999). Averaging over the contributions from mutations arising at each class of fixed sites, taking into account their respective fixation probabilities, the equilibrium rate of nucleotide substitution is then

$$\lambda(\gamma) = \frac{2\kappa\nu\gamma}{[1+\kappa\exp(-\gamma)][\exp(\gamma)-1]} \qquad \text{(A7a)}$$

(Charlesworth and Charlesworth 2010, p.275).

If we consider neutral mutations arising at fixed sites with the same frequencies of $A_1$ and $A_2$ variants as the selected sites (i.e., with the same base composition), the substitution rate is

$$\lambda(0) = \frac{\kappa\nu[1+\exp(-\gamma)]}{[1+\kappa\exp(-\gamma)]} \qquad \text{(A7b)}$$

The ratio $R(\gamma) = \lambda(\gamma)/\lambda(0)$ gives the rate of substitution of selected mutations relative to neutral expectation, conditioning on the same base composition; we have

$$R(\gamma) = \frac{2\gamma}{[\exp(\gamma)-\exp(-\gamma)]} \qquad \text{(A8)}$$

It is easily seen that $R = 1$ at $\gamma = 0$, and decreases as $\gamma$ increases.

**Population distribution statistics for the irreversible mutation model**

The proportion of segregating sites in a sample of size $n$ can be estimated by integrating the probability density of the frequency $p$ of the deleterious allele $A_1$ (Wright 1938), from $1/n$ to $1 - 1/n$. In the present case, this density can be written as

$$\psi(p) = \frac{\alpha[p^{-1} + q^{-1}][\exp - (\gamma p) - \exp(-\gamma)]}{[1 - \exp(-\gamma)]} \qquad (A9)$$

The unconditional population distribution SFS is obtained by multiplying this by $1/n$, similarly to equation (18).

By expanding the integrals of $p^{-1} \exp(-\gamma p)$ and $q^{-1} \exp(-\gamma p)$ as power series, and collecting terms, the proportion of segregating sites is given by

$$\ln(n) + [1 - \exp(-\gamma)] \sum_{i=1}^{\infty} \frac{\gamma^i}{(i+1)i!}[(-1)^i - \exp(-\gamma)] \qquad (A10)$$

Alternatively, the unconditional population distribution SFS can be summed from $1/n$ to $1 - 1/n$; numerical studies show that the two procedures give almost identical results. The conditional SFS is then obtained by normalising the unconditional SFS by the proportion of segregating sites.

**Table 1    Sample statistics for the reversible mutation model ($\kappa = 2$)**

|  | $\beta$ | $p_{seg}$ | $p_{sn}$ | $D_T$ | $\Delta_\pi$ | $p_{seg}$ | $p_{sn}$ | $D_T$ | $\Delta_\pi$ |
|---|---|---|---|---|---|---|---|---|---|
|  |  |  | $n = 20$ |  |  |  | $n = 200$ |  |  |
| $\gamma = 0$ | 0.02 | 0.088 | 0.287 | 0.038 | 0.016 | 0.142 | 0.159 | 0.089 | 0.040 |
|  | 0.2 | 0.533 | 0.219 | 0.322 | 0.109 | 0.728 | 0.086 | 0.745 | 0.326 |
|  | 2 | 0.966 | 0.062 | 1.181 | 0.399 | 1.000 | 0.001 | 1.252 | 1.226 |
|  | 20 | 0.999 | 0.007 | 1.637 | 0.553 | 1.000 | 0.000 | 3.570 | 1.567 |
| $\gamma = 0.5$ | 0.02 | 0.094 | 0.289 | 0.029 | 0.009 | 0.152 | 0.159 | 0.076 | 0.034 |
|  | 0.2 | 0.559 | 0.213 | 0.348 | 0.117 | 0.753 | 0.080 | 0.848 | 0.373 |
|  | 2 | 0.970 | 0.055 | 1.248 | 0.421 | 1.000 | 0.001 | 2.924 | 1.284 |
|  | 20 | 0.999 | 0.007 | 1.649 | 0.556 | 1.000 | 0.000 | 3.586 | 1.574 |
| $\gamma = 5$ | 0.02 | 0.071 | 0.441 | $-0.654$ | $-0.226$ | 0.146 | 0.228 | $-0.843$ | $-0.380$ |
|  | 0.2 | 0.511 | 0.319 | $-0.241$ | $-0.081$ | 0.787 | 0.104 | $-0.032$ | $-0.014$ |
|  | 2 | 0.990 | 0.025 | 0.950 | 0.532 | 1.000 | 0.000 | 3.443 | 1.511 |
|  | 20 | 0.999 | 0.004 | 1.755 | 0.592 | 1.000 | 0.000 | 3.722 | 1.634 |
| $\gamma = 50$ | 0.02 | 0.014 | 0.845 | $-1.539$ | $-0.586$ | 0.063 | 0.478 | $-1.820$ | $-0.851$ |

| | | | | | | | | |
|---|---|---|---|---|---|---|---|---|
| | 0.2 | 0.129 | 0.795 | − 1.650 | − 0.564 | 0.478 | 0.351 | −1.826 | −0.806 |
| | 2 | 0.744 | 0.408 | −0.967 | −0.327 | 0.998 | 0.005 | −0.171 | −0.169 |
| | 20 | 1.000 | 0.000 | 1.329 | 0.736 | 1.000 | 0.000 | 4.270 | 1.875 |

The meanings of the column headings are as follows: $p_{seg}$ is the proportion of sites that are segregating, $p_{sn}$ is the proportion of singletons among segregating sites in a sample of size $n$, $D_T$ is the expected mean Tajima's $D$ for a sequence of 450bp, $\Delta_\pi = (\pi - \theta_w)/\theta_w$, where $\theta_w = p_{seg}/a_n$ and $a_n = 1 + \frac{1}{2} + \ldots 1/(n-1)$. All these statistics were calculated using equations (17).

**Table 2 Comparisons of statistics derived from the sample distribution versus the population distribution**

|  | $p_{seg}$ | | $p_{sn}$ | | $\pi$ | | | | $\theta_w$ | | $D_T$ |
|---|---|---|---|---|---|---|---|---|---|---|---|
|  | Sample | Popn. | Sample | Popn. | Sample | Popn. | Sample | Popn. | Sample | Popn. | |
| $\beta = 0.002$ | | | | | | | | | | | |
| $\gamma = 0$ | 0.0156 | 0.0288 | 0.169 | 0.170 | 0.0026 | 0.0049 | 0.0026 | 0.0049 | 0.007 | 0.007 | |
| $\gamma = 0.5$ | 0.0167 | 0.0311 | 0.171 | 0.172 | 0.0028 | 0.0053 | 0.0028 | 0.0053 | −0.010 | −0.012 | |
| $\gamma = 5$ | 0.0159 | 0.0291 | 0.245 | 0.246 | 0.0016 | 0.0029 | 0.0027 | 0.0050 | −0.752 | −0.832 | |
| $\gamma = 50$ | 0.0068 | 0.0134 | 0.492 | 0.512 | 0.0002 | 0.0003 | 0.0011 | 0.0019 | −1.260 | −1.470 | |
| $\gamma = 500$ | 0.0017 | 0.0006 | 0.849 | 0.958 | 0.0000 | 0.0000 | 0.0002 | 0.0001 | −0.763 | −0.551 | |
| $\beta = 0.02$ | | | | | | | | | | | |
| $\gamma = 0$ | 0.142 | 0.196 | 0.159 | 0.161 | 0.0251 | 0.0345 | 0.0242 | 0.0333 | 0.089 | 0.082 | |
| $\gamma = 0.5$ | 0.152 | 0.209 | 0.159 | 0.161 | 0.0268 | 0.0368 | 0.0259 | 0.0357 | 0.076 | 0.068 | |
| $\gamma = 5$ | 0.146 | 0.201 | 0.228 | 0.232 | 0.0154 | 0.0210 | 0.0249 | 0.0349 | −0.843 | −0.865 | |
| $\gamma = 50$ | 0.063 | 0.083 | 0.478 | 0.501 | 0.0016 | 0.0020 | 0.0107 | 0.0141 | −1.820 | −1.869 | |
| $\gamma = 500$ | 0.014 | 0.005 | 0.844 | 0.957 | 0.0002 | 0.0001 | 0.0023 | 0.0009 | −1.640 | −1.278 | |

The other parameters are: $\kappa = 2$, $n = 200$. See Table 1 for explanation of the meaning of the column headings.

**Figure Legends**

**Figure 1.** The vertical bars are the values (in percentages) of the mean frequency of $A_1$, $\bar{p}$, (red), $\pi$ from equation (11b) (blue), $\pi$ as given by the infinite sites model (black), the proportion of segregating sites from equation (17e) (white), the proportion of segregating sites under the infinite sites model (pink), the 'uncorrected' rate of substitution relative to neutrality (light blue), and the 'corrected' rate of substitution relative to neutrality (green).

**Figure 2.** The curves are the values (in percentages) as functions of $\beta$ for the mean frequency of $A_1$, $\bar{p}$, (red, dashed), $\pi$ from equation (11b) (blue, full), $\pi$ as given by the infinite sites model (green, full), the 'uncorrected' rate of substitution relative to neutrality (black, dashed), and the corrected' rate of substitution relative to neutrality (pink, dashed).

**Figure 3.** The vertical bars are the values (in percentages) of the probabilities of finding the minor allele in a sample of 20 at the frequencies indicated on the $x$ axis, for different values of $\beta$ and $\gamma$.

**Figure 1**

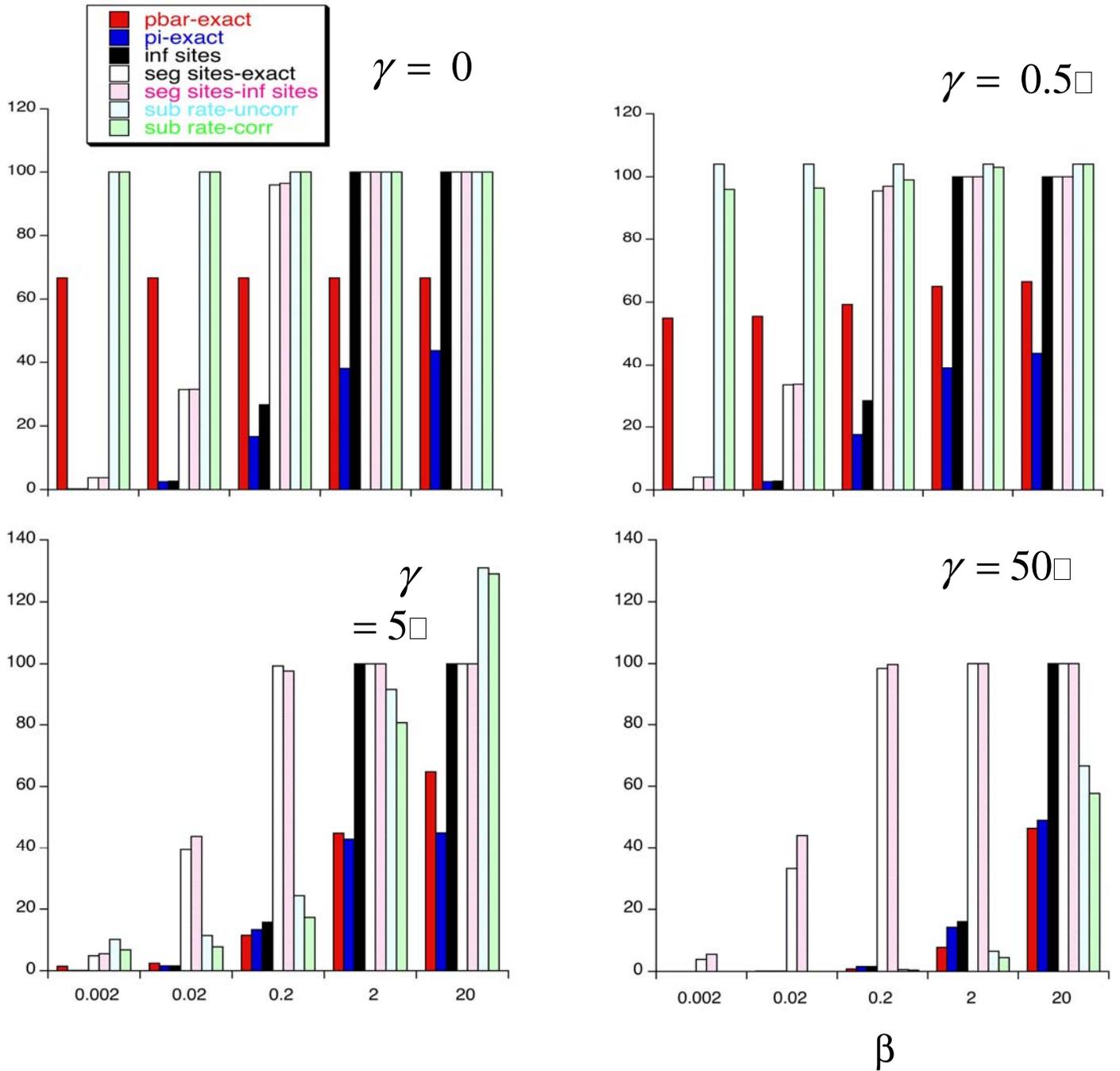

**Figure 2**

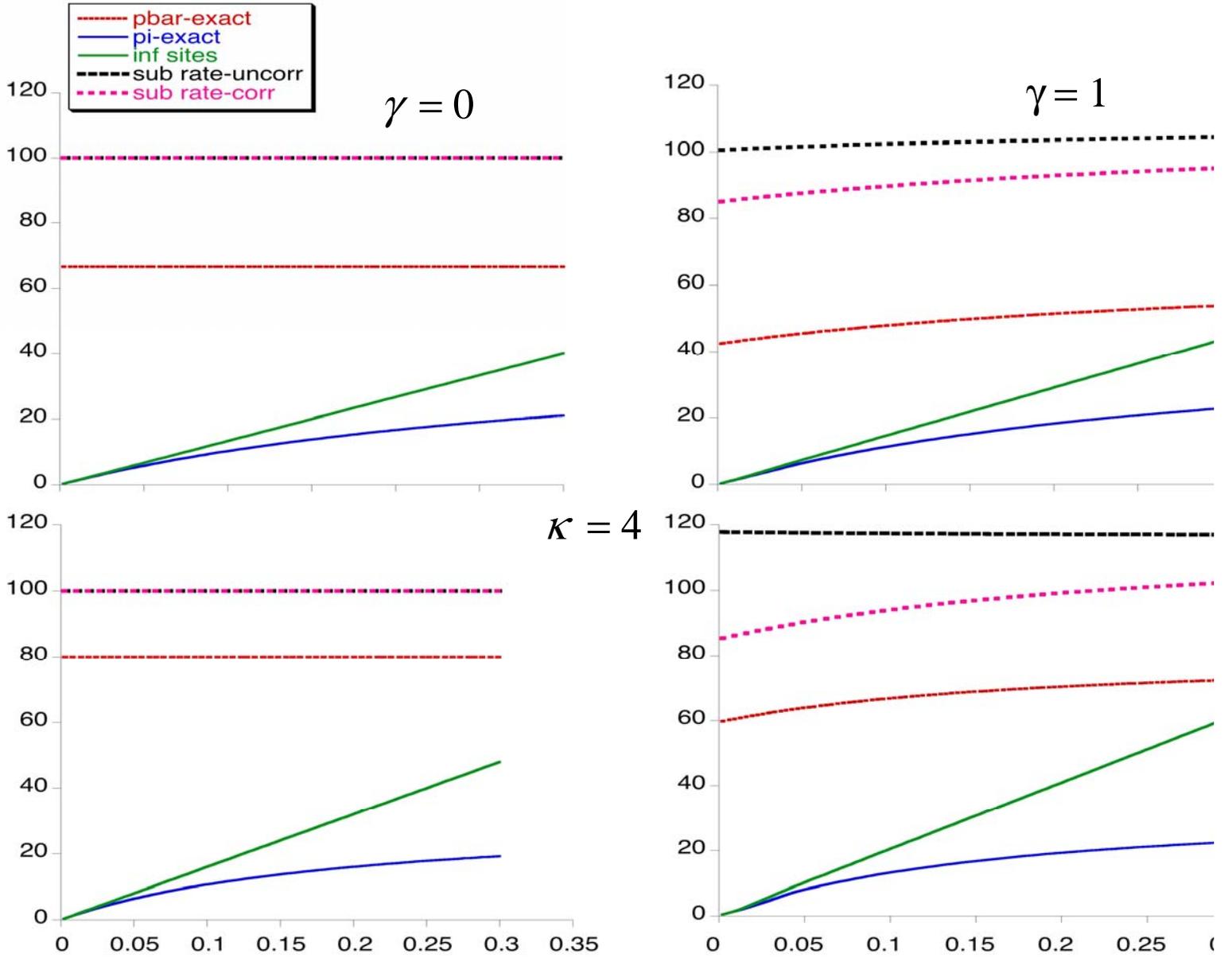

**Figure 3**

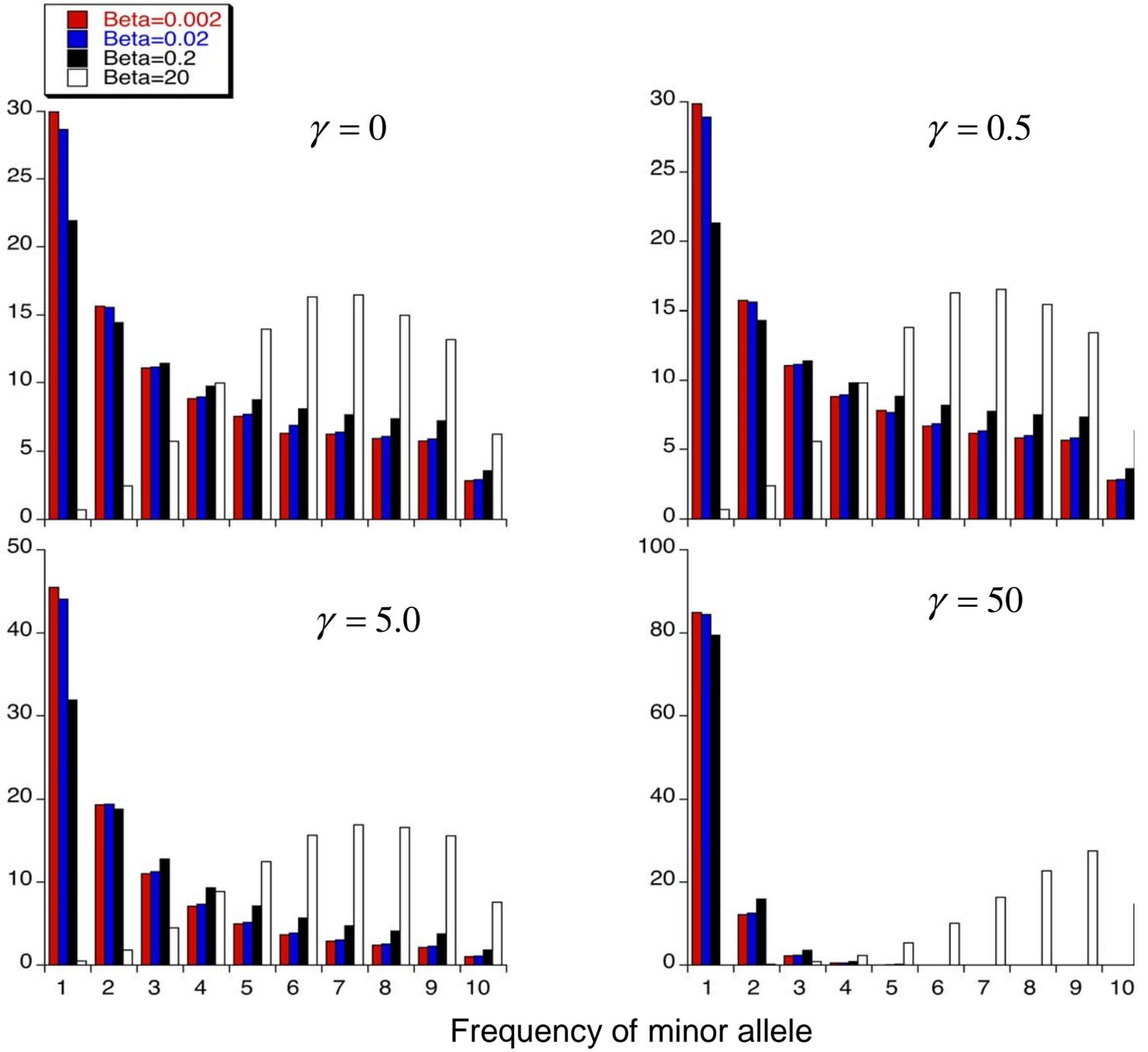

Frequency of minor allele